\documentclass{article}

\def\be{\begin{equation}}
\def\ee{\end{equation}}
\def\bea{\begin{eqnarray}}
\def\eea{\end{eqnarray}}
\def\div{{\rm div}}
\def\curl{{\rm curl}}
\def\d{\partial}
\def\R{{\bf R}}

\newtheorem{theorem}{Theorem}

\begin{document}

\title{Simplified models of electromagnetic and
gravitational radiation damping}

\date{}

\author{{\sc Markus Kunze$^{1}$\& Alan D. Rendall$^{2}$}\\[2ex]
        $^{1}$ Universit\"at Essen, FB 6 -- Mathematik, \\
        D\,-\,45117 Essen, Germany\\
        e-mail: mkunze@ing-math.uni-essen.de \\[1ex]
        $^{2}$ Max-Planck-Institut f\"ur Gravitationsphysik,\\
        Am M\"uhlenberg 1, D\,-\,14476 Golm, Germany \\
        e-mail: rendall@aei-potsdam.mpg.de }

\maketitle

\begin{abstract}\noindent
In previous work the authors analysed the global properties of an
approximate model of radiation damping for charged particles. This
work is put into context and related to the original motivation of
understanding approximations used in the study of gravitational
radiation damping. It is examined to what extent the results
obtained previously depend on the particular model chosen. Comparisons
are made with other models for gravitational and electromagnetic
fields. The relation of the kinetic model for which theorems were
proved to certain many-particle models with radiation damping is
exhibited.
\end{abstract}

\section{Introduction}
\label{intro-sect}

The study of gravitational radiation in general relativity relies in an
essential way on sophisticated approximation methods. It is important to
understand the relation between the approximate models and the exact
theory. Many of the issues arising in this context have recently been
surveyed by Blanchet \cite{blanchet00}. The relation between the exact
and approximate equations is relatively straightforward. What is much
more difficult is to establish a relation between solutions of the two
sets of equations.

A standard approach to the theory of gravitational radiation
(see e.g.~\cite{burke}, \cite{damour87a}, \cite{blanchet00}) is to
combine post-Newtonian and post-Minkowskian approximations via matching.
There are some mathematical results available on these individual types of
approximation. It has been shown that under suitable conditions the
post-Minkowskian approximations are asymptotic \cite{damour90a} but
divergent \cite{rendall90a}. The post-Newtonian approximations are more
difficult to handle. A general mathematical framework for formulating
the relevant questions was set up in \cite{rendall92a}. It was shown
how the well-known divergences of higher order post-Newtonian
approximations could be understood within that approach. This
framework also provided a basis for showing that the post-Newtonian
expansion is asymptotic at the lowest order (Newtonian) level
\cite{rendall94a}. It is reasonable to expect that analogous results
could be proved for the 1PN and 2PN levels but this has not been
done yet.

None of the results mentioned above prove anything about matching
the two approximations and until this can be done the possibilities
of understanding anything about radiation (even at the quadrupole
level) in a rigorous way are very limited. In the following these
limits will not be removed. A more modest goal is pursued, which is
to understand some of the effective models for radiation
themselves while postponing the question of their relation to the
exact equations. Radiation comes in at the 2.5PN level, but there
are hybrid models where radiation at that level is combined with
a description of matter and non-radiative gravitational fields at
the Newtonian or 1PN level. These models will be referred to as
(0+2.5)PN and (1+2.5)PN models.

For purposes of comparison it is valuable to consider the case of
charged matter emitting electromagnetic radiation as an analogue
of matter emitting gravitational radiation. In this case the
post-Minkowskian expansion becomes trivial due to the linearity
of the Maxwell field while there is a non-trivial analogue of the
post-Newtonian expansion. In the electromagnetic case radiation
first appears at the 1.5PN level and it is possible to define
(0+1.5)PN and (1+1.5)PN models. These will be referred to as the
Coulomb and Darwin cases respectively.

In \cite{kunze00a} the authors proved theorems on the global existence and
asymptotic behaviour of solutions of a certain model of the
(0+1.5)PN type in the electromagnetic case. The purpose of the present
paper is to put this mathematical result into a wider context and to
discuss its ramifications. For technical details of the proof
the reader is referred to \cite{kunze00a}.

A general mathematical setting for expansions of PN type is as follows.
Matter is described by a function $f(\lambda)$ depending on a parameter
$\lambda$ belonging to an interval $(0,\lambda_0]$. The field
(gravitational or electromagnetic) is described by a function
$g(\lambda)$. The idea is to consider a one-parameter family of
solutions which become non-relativistic in the limit $\lambda\to 0$
in some appropriate sense. The parameter $\lambda$ corresponds to
$c^{-2}$ where $c$ is the speed of light. More precisely, the given
functions $f$ and $g$ describe a one-parameter family of physical
systems represented in suitable parameter-dependent units. Then $c$ is
to be interpreted as the numerical value of the speed of light in the
given units. A more conventional physical description of the
post-Newtonian expansion would say, for instance, that velocities of
the matter are small with respect to the speed of light. The connection
between the two descriptions is that to say that the matter velocity is
of order unity and the speed of light large in one system of units is
equivalent to saying that the matter velocity is small and the speed of
light has a fixed value in a different (fixed) system of units.

In our discussion of expansions for charged matter we will start with the
following formal expressions:
\bea
E&=&E_0+\lambda^{1/2} E_1+\lambda E_2+\lambda^{3/2}E_3+\ldots\,,
\label{E-expa} \\
B&=&B_0+\lambda^{1/2}B_1+\lambda B_2+\lambda^{3/2}B_3+\ldots\,,
\label{B-expa} \\
\rho&=&\rho_0+\lambda^{1/2}\rho_1+\lambda\rho_2+\lambda^{3/2}\rho_3
+\ldots\,, \\
j&=&j_0+\lambda^{1/2}j_1+\lambda j_2+\lambda^{3/2}j_3+\ldots\,.
\eea
The use of expansions in half-integral powers of a parameter $\lambda$
ensures that the notation fits with the usual terminology of post-Newtonian
expansions. If we write
$\lambda=c^{-2}$ then the Maxwell equations are:
\bea
&\div E=4\pi\rho, &\curl E=-\lambda^{1/2}\d_t B,  \label{E-eq} \\
&\div B=0,  &\curl B=\lambda^{1/2}\d_t E+4\pi\lambda^{1/2} j. \label{B-eq}
\eea
Substituting the expansions into the equations and comparing coefficients
gives a sequence of equations for these coefficients. Note that $\div B_0=0$
and $\curl B_0=0$. It follows that assuming fields defined on $\R^3$ and
vanishing at infinity $B_0=0$. This condition will be assumed from now on.
The first equations for coefficients of $E$ of the sequence (ordering by
powers of $\lambda$) are
$\div E_0=4\pi\rho_0$ and $\curl E_0=0$. The second equation implies that
$E_0$ is the gradient of a function $U$ and then the first equation implies
that $\Delta U=4\pi\rho_0$. Thus the Poisson equation of electrostatics is
recovered, with $U$ being the electrostatic potential.

Next a remark will be made concerning the coefficients of the odd half-integer
powers in the expansion of $E$, $\rho$ and $j$ and the integer powers in the
expansion of $B$. These can be consistently set to zero. In fact if the
coefficients belonging to this class are set to zero in the matter quantities
then (assuming fields defined on $\R^3$ and vanishing at infinity) the
coefficients of the fields belonging to this class automatically vanish. In
this section it will be assumed that all these coefficients are identically
zero. They will, however, be important in the later section \ref{Ntoinfty}.
In a pure post-Newtonian expansion like the above they play no essential role
and if they are maintained they just lead to \lq shadow equations\rq\ as
discussed in \cite{rendall92a}.

Consider next $B_1$. It satisfies the equations $\div B_1=0$
and $\curl B_1=\d_t E_0+4\pi j_0$. Combining these equations and using the fact
that $\curl E_0=\div B_1=0$ gives $\Delta B_1=-4\pi\curl j_0$.
Thus $B_1$ solves a Poisson equation with a source determined by the
lowest order coefficient in the expansion of the current. It is easily
shown, using the asymptotic conditions assumed above, that the
solution of this Poisson equation satisfies the original first order
equations. Next combining the equations $\div E_2=4\pi\rho_2$
and $\curl E_2=-\partial_t B_1$ and substituting for $\curl B_1$ leads to the equation
\be
\Delta E_2=\d_t^2 E_0+4\pi(\nabla\rho_2+\d_t j_0).
\ee
Assuming that the matter variables have compact support for each fixed value
of $t$ the second term on the right hand side of this equation also has
compact support and leads to no difficulties in solving this Poisson
equation. It remains to discuss the first term, which is not compactly
supported. Assume that the matter sources satisfy the continuity equation
$\d_t\rho+\div j=0$. Then the total charge, $\int \rho (x) dx$, is
independent of time and the first term on the right hand side of the
Poisson equation for $E_2$ is $O(r^{-3})$. It follows (for example from
results in the appendix of \cite{rendall92a}) that there is a solution for
$E_2$ which vanishes at infinity and is unique in a suitable function space.
The case of $B_3$ is more problematic. According to the assumptions made up
to now it should satisfy the Poisson equation:
\be
\Delta B_3=\d_t^2 B_1-4\pi\curl j_2.
\ee
Once again the second term is unproblematic, but the first needs to be looked
at more closely since it might have a contribution which only falls off like
$r^{-1}$ as $r\to\infty$. This contribution is made by the second time
derivative of the quantity $\int j_0(x) dx$. Using the continuity equation
and partial integration the latter quantity can be rewritten as
$-(d/dt)\int x\rho_0 (x) dx$. This is (up to sign) the first time derivative
of the dipole moment $D(t)$ of the charge distribution at the Coulomb level.
Thus the coefficient which is problematic
is $d^3 D/dt^3$ and this is a quantity which comes up in the description of
radiation damping. What has happened here, roughly speaking, is that with
$B_3$ which is the coefficient in a term of order $\lambda^{3/2}$ we have
reached the level at which
radiation damping plays a role and the naive post-Newtonian expansion breaks
down. A correct analysis using matching would show that the asymptotic
conditions to be imposed on $B_3$ are different from the vanishing of that
quantity at infinity. With a right hand side of the given form the Poisson
integral will diverge and it is not to be expected that the Poisson equation
for $B_3$ will have a solution which vanishes at infinity. In any case it
does not have a solution in the kind of weighted spaces introduced in
\cite{rendall92a}. Thus the expansion here breaks down at the 1.5PN
level in much the same way as the post-Newtonian expansion for gravity
breaks down at the 3PN level (as shown in \cite{rendall92a}).

Adopting terminology adapted to that used in \cite{degond92} we call a
solution of the equations for $E_0$ the Coulomb approximation, a
solution of the equations for $E_0$ and $B_1$ the quasi-electrostatic
approximation, and a solution of the equations for $E_0$, $B_1$ and $E_2$
the Darwin approximation.

In order to throw some more light on the particular form of the expansions
chosen it is useful to reexpress the Maxwell equations in four-dimensional
form. The Maxwell equations in Minkowski space are
$\d_\alpha F^\alpha{}_\beta=4\pi j_\beta$ and
$\d_\alpha F_{\beta\gamma}+\d_\gamma F_{\alpha\beta}
+\d_\beta F_{\gamma\alpha}=0$. Here $F^{\alpha\beta}$ is the field tensor
and $j^\alpha$ is the four-current. The expansions above can be written
in terms of one-parameter families of these quantities. Since we are
interested in the Newtonian limit it is useful to introduce a family
of metrics $f_{\alpha\beta}(\lambda)$ as in \cite{rendall92a} by $f_{00}=-1$,
$f_{0a}=0$, $f_{ab}=\lambda\delta_{ab}$. All these metrics are flat
and in fact represent the Minkowski metric in different units. As basic
quantities we take one-parameter families $F^\alpha{}_\beta(\lambda)$ and
$j^\alpha (\lambda)$ defined for $\lambda>0$ and assume that they have
limits for $\lambda\to 0$. The tensor $f_{\alpha\beta}(\lambda)$ also
has a limit for $\lambda\to 0$ but the limit is not a Lorentz metric. In
general indices will be raised and lowered using the metric $f_{\alpha\beta}$
and its inverse.

Define $E^a=F^a{}_0$, $B^1=\lambda^{-1/2} F^2{}_3$,
$B^2=\lambda^{-1/2} F^3{}_1$, $B^3=\lambda^{-1/2} F^1{}_2$. The last three
relations can also be expressed
in terms of the spatial volume form associated to $f$, call it $\epsilon_f$.
Using this we can write $B^a=(\epsilon_f)^a{}_b{}^c F^b{}_c$. In general
situations will be considered where $F^a{}_0$ has a non-zero limit and then
the same is true of $E^a$. On the other hand it will be assumed that $F^a{}_b$
is $O(\lambda)$ as $\lambda\to 0$ and then $B^a$ is
automatically $O(\lambda^{1/2})$. Define $\rho=j^0$ and take the current in
the three-dimensional form of the Maxwell equations to have components $j^a$.
With these definitions the three- and four-dimensional forms of the Maxwell
equations are equivalent and the expansions of $E$, $B$ and $j$ above
correspond to expanding $F^\alpha{}_\beta$ in integral powers of $\lambda$,
starting with the power zero, and assuming that the first coefficient in
the expansion is zero when both $\alpha$ and $\beta$ are spatial. The
correctness of this statement depends on the fact that the coefficients of
the odd half-integer powers in the expansion of $E$, $\rho$ and $j$ and the
integer powers in the expansion of $B$ have been set to zero.
Radiation damping comes in at order $\lambda^{3/2}$ and
this is not allowed for by the expansion in integral powers of $\lambda$.
However as long as we are doing an expansion of post-Newtonian type it
is the case that if $j^\alpha$ contains only integral powers of $\lambda$
the same is true of $F^\alpha{}_\beta$.

In the practical applications of post-Newtonian approximations it is usual
to expand the fields but not the matter quantities. This point has been
discussed briefly in section 6 of \cite{rendall92a}. What this means in
effect is that we are not talking about functions which solve a certain
system of equations exactly but rather about parameter-dependent families
of functions which satisfy certain equations up to an error of the order
of a power of the parameter. Related to this is the fact that the equations
themselves are often only fixed up to an error of a certain order. In reality
we are dealing with equivalence classes of equations.  This point will be
illustrated by comparing the Darwin model introduced by Degond and Raviart in
\cite{degond92} with the Darwin approximation introduced above. Degond and
Raviart split the electric field into transverse and longitudinal parts
and set the transverse part of $E$ to zero in the Maxwell equation
containing $\curl B$ to obtain the Darwin model. We claim that at the level
of what we call the Darwin approximation the expansions of the Darwin model
of \cite{degond92} and the full Maxwell equations are identical. Thus at
that level both sets of equations are in the same equivalence class. In
terms of our expansion the transverse part of $E$ is that arising from
the equations for $\curl E_n$ while the equation containing $\curl B$
gives rise to the equations for $\curl B_n$. The only coefficient $B_n$
which occurs at the level of the Darwin approximation is $B_1$, in the
equation $\curl B_1$ the only contribution from the electric field comes
from $E_0$, and $E_0$ has no transverse part. Thus on this level the passage
from the full Maxwell equations to the Darwin model has no effect.

Up to this point we have concentrated on the fields and said little about
the matter. The equations of motion for charged particles contain the
Lorentz force $F_L=E+\lambda^{1/2}v\times B$. In terms of the expansion
coefficients this gives $E_0+\lambda(E_2+v\times B_1)$ as the contribution
up to order $\lambda$. The most common choices of matter models when
discussing approximation schemes are point particles or a perfect fluid.
In general relativity the concept of a point particle is very problematic
and can itself only be defined in an approximate sense. A perfect fluid
is better, but when it comes to obtaining mathematically rigorous results
it also has its problems. On the one hand there is the problem of the formation
of shocks, which is always a danger when considering solutions of the
Euler equations on a long time scale. When dealing with radiation we do
wish to consider the evolution of the system on a long time scale. Once
shocks are formed we enter a regime where little control is possible with
known mathematical techniques.  The other problem concerns the question of
fluid bodies. In studying radiation we want to consider isolated systems.
The easiest thing would be to take the fluid to have compact support.
Unfortunately the Cauchy problem for solutions of the Euler equations with
compactly supported initial density is poorly understood, even locally in
time. There has been some progress, since Wu \cite{wu99} proved a local
existence theorem for the incompressible Euler equations with free boundary
in the irrotational case. Christodoulou and Lindblad \cite{christ00} proved
estimates for the case with rotation which will hopefully lead to an
analogous result in that situation. However much remains to be understood in
that area. It might be thought that replacing compactly supported initial
data with data which fall off at infinity might simplify the Cauchy problem
for a fluid body but nobody has managed to take advantage of this yet.

In Section 2 models of radiation which combine expansions at PN levels which
belong to different powers of the expansion parameter
(\lq hybrid models\rq ) are discussed. After reviewing the results of
\cite{kunze00a} on one model of this type for electromagnetic radiation we
discuss various issues relating to the possibility of generalizing those
results to related models for electromagnetism and gravitation. In the model
of \cite{kunze00a} the matter is described by kinetic theory. Section 3
of this paper relates the kinetic model of \cite{kunze00a} to the underlying
dynamics of a many-particle system with radiation reaction, thus providing
a better justification for the kinetic model which was originally introduced
in an ad hoc manner. The paper concludes with a brief outlook.

\section{Hybrid models}
\label{hyb-sect}

By a hybrid model we mean a model where an expansion of post-Newtonian
type up to a certain order is combined with effects of radiation which
are formally of higher order in the expansion parameter $\lambda$. Some
variants of this procedure were already mentioned in the introduction.
The model studied in \cite{kunze00a} was for the case of the electromagnetic
field with a kinetic description of the matter at the Newtonian level and
dipole radiation at the 1.5PN level. It was inspired by a model of
Blanchet, Damour and Sch\"afer \cite{bds} for the case of the gravitational
field with a hydrodynamic description of matter at either the Newtonian or
1PN level and quadrupole radiation at the 2.5PN level. The replacement of
the fluid by kinetic theory was motivated by the difficulties with a
mathematical treatment of fluids mentioned in the introduction. The
replacement of the gravitational by the electromagnetic field was
motivated by simplicity. The extent to which the particular choice
of model in \cite{kunze00a} was essential in obtaining results will be
discussed below.

In \cite{bds} an initial model with higher order (in fact fifth order) time
derivatives which could not be expected to have a good initial value
formulation was modified by two methods for reducing the order of
the time derivatives. The resulting model is not equivalent to the original
one but is in the same equivalence class in a sense already indicated.
In other words the equations of the two models agree up to terms which
are regarded as being of higher order at the given level of approximation.
One method of reduction is to do an appropriate change of matter variables.
The other is to substitute the equations of motion into the undesirable
time derivatives and to discard terms which arise from this which are
formally of higher order. In \cite{bds} the first method is used to remove
two time derivatives and the second to remove the remaining three. Thus
we could call this a (0+2.5)PN or (1+2.5)PN model with 3+2 reduction of time
derivatives. The model of \cite{kunze00a} is a (0+1.5)PN model with 2+1
reduction of time derivatives.

The results of \cite{kunze00a} will now be stated. For reasons explained in
that paper it is necessary to have species of particles with different
charge to mass ratios in order to get an interesting result. The particular
choice made was to have two species of particles with unit mass and charges
of unit magnitude and opposite sign. The phase space densities of these
two species of particles are described by functions $f^+$ and $f^-$. These
functions satisfy the Vlasov equation with a force term which is the sum
of a Coulomb term resulting from a potential $U$ and a radiation reaction
term. The primitive form of the latter term is (up to sign)
$\epsilon d^3 D/dt^3$ where
\begin{equation}\label{D-def}
   D(t)=\int x\,(\rho^+(t, x)-\rho^-(t, x))\,dx
\end{equation}
is the dipole moment of the charge distribution and $\epsilon$ is a small constant.
The charge distribution itself is defined to be
\be
\rho(t,x)=\rho^+(t,x)-\rho^-(t,x)=\int f^+(t,x,v)-f^-(t,x,v)\,dv.
\ee
The primitive form is reduced by procedures of the type already mentioned in
such a way that the equations for $f^+$ and $f^-$ are of the form
\bea
&\d_t f^++(v+\epsilon D^{[2]}(t))\cdot\nabla_x f^++\nabla U\cdot\nabla_v f^+=0,
\label{vlas+}
\\
&\d_t f^-+(v-\epsilon D^{[2]}(t))\cdot\nabla_x f^--\nabla U\cdot\nabla_v f^-=0,
\label{vlas-}
\eea
where
\be\label{D2-def}
D^{[2]}(t)=\int \nabla U(t,x)(\rho^+(t, x)+\rho^-(t, x))\,dx.
\ee
The full system of equations for $U$ and $f$ is called the VPD system
(Vlasov-Poisson with damping). Define the total energy $\cal E$ of the
system by
\begin{equation}\label{etot-def}
   {\cal E}(t)={\cal E}_{{\rm kin}}(t)+{\cal E}_{{\rm pot}}(t),
\end{equation}
with
\begin{eqnarray}
   {\cal E}_{{\rm kin}}(t) & =& \frac{1}{2}\int\int |v|^2 (f^+(t, x, v)
   +f^-(t, x, v))\,dx dv,\quad\mbox{and} \nonumber \\
   {\cal E}_{{\rm pot}}(t)
   & =& \frac{1}{8\pi}\int |\nabla U(t, x)|^2\,dx,
\label{ekin-def}
\end{eqnarray}
denoting kinetic and potential energy, respectively. With these
preliminaries in hand we can state the main theorem of \cite{kunze00a}.

\begin{theorem}
If $f^{\pm}_0$ are smooth initial data with compact support for the VPD
system, then there is a unique smooth solution $f^{\pm}$ of VPD for
$t\ge 0$ with data $f^{\pm}(t=0)=f^{\pm}_0$. The energy evolves according
to
\begin{equation}\label{energ-bd}
   \dot{{\cal E}}(t) = -\epsilon\,{|D^{[2]}(t)|}^2.
\end{equation}
Moreover, the following estimates hold for $t\in [0, \infty[$.
\begin{itemize}
\item[(a)] $\displaystyle {\|\rho^{\pm}(t)\|}_p
\le C{(1+t)}^{-\frac{3(p-1)}{2p}}$ for $p\in [1, \frac{5}{3}]$;
\item[(b)] $\displaystyle {\|\nabla U(t)\|}_p
\le C{(1+t)}^{-\frac{5p-3}{7p}}$ for $p\in [2, \frac{15}{4}]$;
\item[(c)] $|D^{[2]}(t)|\le C{(1+t)}^{-\frac{8}{7}}$.
\end{itemize}
Here $\|\ \|_p$ denotes the $L^p$ norm.
\end{theorem}
Thus we obtain a global in time existence theorem and statements about
how various quantities decay as $t\to\infty$. The theorem was proved by
adapting methods previously applied to the usual Vlasov-Poisson system by
Lions and Perthame \cite{lions91} (global existence) and Illner and Rein
\cite{illner96} (asymptotic decay).

Next we consider whether similar results can be obtained for other hybrid
models. First we give the equations for some of these models. Starting
with the primitive form of the equations for charged particles described
by the Vlasov equation at the (0+1.5)PN level, a 1+2 reduction
can be carried out. This means introducing new variables in the
characteristic equations by $\tilde X^{\pm}=X^{\pm}\mp\epsilon\dot D(t)$ and
$\tilde V^{\pm}=V^{\pm}\mp\epsilon\ddot D(t)$ with the result that the new
characteristic equations are $\dot{\tilde X}^{\pm}=\tilde V^{\pm}$ and
$\dot{\tilde V}^{\pm}=-\nabla U (t,\tilde X^{\pm}\pm\epsilon\dot D(t))$.
The Vlasov equation with these characteristics has the form
\be
\d_t f^{\pm}+v\cdot\nabla_x f^{\pm}
-\nabla U (t,x\pm\epsilon \dot D(t))\cdot\nabla_v f^{\pm}=0.
\ee
Then the other method must be applied to replace $\dot D(t)$ by $D^{[1]}(t)$,
where
\[ D^{[1]}(t)=\int\int v(f^+-f^-)dx dv, \]
cf.~section 1 of \cite{kunze00a}. This defines the model.
There is no obvious obstacle to proving local existence for this case.
The other hybrid model obtained from the same starting point
is that resulting from a 3+0 reduction. In that case the Vlasov equation
is of the form
\be
\d_t f^{\pm}+v\cdot\nabla_x f^{\pm}
\pm (\nabla U (t,x)+\epsilon D^{[3]}(t))\cdot
\nabla_v f^{\pm}=0
\ee
where
\be
D^{[3]}(t)=\int\int \nabla\dot U(f^++f^-)dx dv+\int\int \nabla\nabla U\cdot
v (f^++f^-) dx dv.
\ee
In this case proving local existence may be more difficult, due to the second
derivatives of $U$ occurring in the Vlasov equation. There is a loss of one
derivative compared to the 2+1 and 1+2 reductions.

The consequences of the lost derivative will now be examined in some more
detail. In order to get a local existence theorem the orders of
differentiability of the different unknowns must fit together properly.
Suppose that $U$ is $k$ times differentiable. Then $f$, and hence $\rho$
will be $k-2$ times differentiable. Thus, in order that a match be obtained,
two derivatives must be gained when solving the Poisson equation. It is well
known (see e.g.~\cite{gilbarg83}) that this is not true in the context of
pointwise differentiability properties. It is true when regularity is
measured in H\"older or Sobolev spaces \cite{gilbarg83}. Thus in order to
get a local existence theorem it is either necessary
to work with spaces of that kind or to see that at some point a derivative
can be gained in comparison to the naive counting just given. This appears as
a borderline case for local existence.

In the case of the 2+1 reduction studied in \cite{kunze00a} a simple and
convenient formula for the energy loss of the system due to emission of
radiation was obtained. For the 3+0 and 1+2 reductions things seem to
be more complicated. In the model considered in \cite{kunze00a} the rate
of change of the total energy is $-\epsilon |D^{[2]}(t)|^2$, which is
manifestly negative. The corresponding expression for the 3+0 reduction
is $\epsilon D^{[1]}(t)\cdot D^{[3]}(t)$, which is neither manifestly negative
nor obviously related to the other expression. On a formal level one link
can be seen. Suppose that we replace the quantities $D^{[n]}(t)$ by
the true derivatives $D^{(n)}(t)$. Then we have the identity that
\be
D^{(1)}(t)\cdot D^{(3)}(t)=-|D^{(2)}(t)|^2+\d_t(D^{(1)}(t)\cdot D^{(2)}(t)).
\ee
If we suppose that the function $D(t)$ has a suitable oscillatory
behaviour then the last term will average to zero so that in an average
sense the left hand side is equal to the first term on the right hand
side. To turn this formal argument into a rigorous one would require
at the very least a lot more information than was obtained in
\cite{kunze00a}. The relation to the energy loss in the 1+2 reduction is
even less clear.

Consider next a model with gravitation and kinetic theory at the (0+2.5)PN
level. In comparison to the electromagnetic case the main difference is that
the third derivative of the dipole moment is replaced by the fifth derivative
of the quadrupole moment. The primitive form of the radiation reaction
term is (up to sign) $\epsilon (d^5 Q_{ab}/d t^5)x^a$, where the quadrupole
moment is given by
\be
Q_{ab}(t)=\int \Big(x_ax_b-\frac{1}{3}|x|^2\delta_{ab}\Big)\rho (t, x)\,dx.
\ee
The fifth order derivative is reduced to the third order derivative by
the transformation (cf.~equation (10) in \cite{kunze00a})
\bea
&\tilde X=X-\epsilon Q^{(3)}(t)\cdot X,             \\
&\tilde V=V-\epsilon Q^{(4)}(t)\cdot X+\epsilon Q^{(3)}(t)\cdot V.
\eea
The transformed Vlasov equation is
\bea
&\d_t f+(v-2\epsilon Q^{(3)}(t)\cdot v)\cdot\nabla_x f \nonumber\\
&-[\nabla U(t,x+\epsilon Q^{(3)}(t))
-\epsilon Q^{(3)}(t)\cdot\nabla U(t,x)]\cdot\nabla_v f=0.
\eea
In doing this transformation terms which are formally of order $\epsilon^2$
have been discarded.
The next step in the 3+2 reduction of the gravitational case is to replace
the time derivative $Q^{(3)}$ by the reduced quantity $Q^{[3]}$. This gives
rise to second derivatives of the potential as in the 3+0 reduction of the
electromagnetic case. To
write the expression for the reduced time derivative $Q^{[3]}$ in a compact
form it is useful to introduce the following operation on tensors
\be
{\rm STF}(T_{ab})=(1/2)(T_{ab}+T_{ba})-(1/3)\delta^{cd}T_{cd}\delta_{ab}.
\ee
It takes the symmetric trace-free part of a given tensor. Then
\be
Q^{[3]}(t)={\rm STF}\bigg(2\int\int[(v\cdot\nabla\nabla U+\nabla\dot U)\otimes x
+3\nabla U\otimes v ] f(t,x,v)\,dx dv\bigg).
\ee
This may be compared with equation (6.6) of \cite{bds} where a
corresponding equation is given for the hydrodynamic case. Two of
the terms above can easily be seen to correspond to terms in the
equation of \cite{bds}. The sum of the remaining terms in the latter
equation correspond to the term above involving $\dot U$ via the
equations of motion. There does not seem to be a useful way to
eliminate the time derivative of the potential in the case where the
matter is described by kinetic theory.

As in the 3+0 reduction in the electromagnetic case the second derivatives
may make proving local existence difficult. Note that a 4+1 reduction
would lead to the appearance of third derivatives of the potential, the loss
of two derivatives compared the model treated in \cite{kunze00a} and,
presumably, to ill-posedness of the evolution equations.

\section{Comparison with $N$-particle systems}\label{Ntoinfty}

The VPD model considered above has been introduced as an effective model
for electromagnetic radiation. The purpose of this section is to see how this
model relates to the effective equations for $N$-particle systems
of Abraham-Lorentz type \cite{MK-S, MK-S-2, Sbook} in the limit $N\to\infty$.
As before we assume the fields $E$ and $B$ do satisfy the Maxwell equations
(\ref{E-eq}), (\ref{B-eq}), and they are expanded as in (\ref{E-expa}),
(\ref{B-expa}) in $\lambda^{1/2}=c^{-1}$. Contrary to section \ref{intro-sect},
however, we do not suppose a priori that the coefficients of the odd
half-integer
powers in the expansion of $E$, $\rho$ and $j$ and the integer powers in the
expansion of $B$ are set to zero.

\subsection{The Darwin approximation}
\label{CD-sect}

In section \ref{intro-sect} we have already seen that the first few equations
the coefficients have to satisfy are $B_0=0$ and
\begin{eqnarray*}
   & & \Delta E_0=4\pi\nabla\rho_0,\quad\Delta B_1=-4\pi\,\curl j_0, \\
   & & \Delta E_1=4\pi\nabla\rho_1,\quad\Delta E_2=4\pi\nabla\rho_2
   +\partial_t(\partial_t E_0+4\pi j_0).
\end{eqnarray*}
We intend to derive an explicit approximation of the Lorentz force
$F_L=E+\lambda^{1/2}v\times B$ up to order ${\cal O}(\lambda)
={\cal O}(c^{-2})$, given through
\begin{equation}\label{Lor-force}
   F_L=E_0+\lambda^{1/2}E_1+\lambda (E_2+v\times B_1)
   +{\cal O}(\lambda^{3/2}).
\end{equation}
Suppressing the time argument, clearly
\begin{equation}\label{E0-form}
   E_0(x)=4\pi\Delta^{-1}(\nabla\rho_0)(x)
   =\int\frac{x-y}{|x-y|^3}\,\rho_0(y)\,dy,
\end{equation}
with the analogous formula for $E_1$ by replacing $\rho_0$ by $\rho_1$.
Moreover,
\begin{eqnarray}\label{B1-form}
   B_1(x) & = & -4\pi\Delta^{-1}(\curl j_0)(x)
   =-\int\frac{x-y}{|x-y|^3}\times j_0(y)\,dy
   \nonumber\\ & = & -\int\frac{x-y}{|x-y|^3}\times j(y)\,dy
   +{\cal O}(\lambda^{1/2}).
\end{eqnarray}
To determine $E_2$, we note that $\partial_t\rho_0
+{\rm div}j_0=0$ in view of $\div E_0=4\pi\rho_0$
and $\curl B_1=\partial_t E_0+4\pi j_0$, whence
\[ \Delta(\partial_t E_0)=4\pi\nabla(\partial_t\rho_0)
   =-4\pi\nabla({\rm div}j_0). \]
Accordingly, $E_2$ satisfies
\begin{eqnarray*}
   E_2 & = & 4\pi\Delta^{-1}(\nabla\rho_2)+\partial_t\,\Delta^{-1}
   \Big(\partial_t E_0+4\pi j_0\Big) \\ & = & 4\pi\Delta^{-1}(\nabla\rho_2)
   +4\pi\partial_t\,\Big(-\Delta^{-2}(\nabla {\rm div}j_0)+\Delta^{-1}j_0\Big).
\end{eqnarray*}
Here we follow the device of \cite[p.~296]{bds} how to solve iterated
Poisson equations, and observing that $X(x)=\int |x-y|\,\sigma(y)\,d^3y$
is the solution to $\Delta^2 X=-8\pi\sigma$, after some calculation
the more explicit form
\begin{eqnarray}\label{E2-form}
   E_2(x) & = & 4\pi\Delta^{-1}(\nabla\rho_2)(x)
   -\frac{1}{2}\,\partial_t\,\bigg(\int\frac{1}{|x-y|}\,j_0(y)\,dy
   \nonumber \\ & & \hspace{5em}
   +\int\frac{1}{|x-y|^3}\,(x-y)\cdot j_0(y)\,(x-y)\,dy\bigg)
\end{eqnarray}
is found. Conversely, it may be verified directly that this function $E_2$
obeys ${\rm div}E_2=4\pi\rho_2$ and $\curl E_2=-\partial_t B_1$.
Taking into account (\ref{E0-form}), (\ref{B1-form}), and (\ref{E2-form}),
we have obtained an explicit representation for $F_L$ from (\ref{Lor-force})
up to order ${\cal O}(\lambda)$.

To make the connection with the $N\to\infty$ limit of a system
of $N$ particles we need to rewrite $E_2$ further and make use
of the fact that this description is not unique since terms
of order ${\cal O}(\lambda^{3/2})$ can be added or subtracted.
Adopting the terminology from section \ref{intro-sect} this means
that we look for a more convenient representative
in the same equivalence class of equations. {}From now on we assume
that $f=f_0+\lambda^{1/2}f_1+\lambda f_2+\ldots$ satisfies
the relativistic Vlasov equation
\begin{equation}\label{vlas}
   \partial_t f+v\cdot\nabla_x f+(E+\lambda^{1/2}\,v\times B)\cdot
   \nabla_p f=0
\end{equation}
coupled to the Maxwell equations (\ref{E-eq}), (\ref{B-eq}).
Here $p\in\R^3$ denotes the momentum, thus $v=(1+\lambda p^2)^{-1/2}p$
for the velocity if we consider particles of unit mass.
Therefore $v=p+{\cal O}(\lambda)$ implies that
\begin{equation}\label{vlasov-poisson}
   0=\partial_t f_0+p\cdot\nabla_x f_0
   +E_0\cdot\nabla_p f_0+{\cal O}(\lambda^{1/2})
\end{equation}
i.e., to lowest order $f\cong f_0$ is a solution of the usual Vlasov-Poisson system
with electric field $E_0$ obeying $\Delta E_0=4\pi\nabla\rho_0$.
Hence we can invoke (\ref{vlasov-poisson}) and $j_0=\int vf_0\,dp$ to write
\begin{eqnarray}\label{dtj0}
  \partial_t j_0 & = & -\int p\,(p\cdot\nabla_x f_0
   +E_0\cdot\nabla_p f_0)\,dp+{\cal O}(\lambda^{1/2})
   \nonumber \\ & = & -\int p(p\cdot\nabla_x f_0)\,dp
   +\int E_0 f_0\,dp+{\cal O}(\lambda^{1/2}).
\end{eqnarray}
Next we explicitly perform the $\partial_t$-differentiation
in (\ref{E2-form}) and substitute the approximate expression
for $\partial_t j_0$ from (\ref{dtj0}). We obtain after some simplification,
including $E_0=4\pi\Delta^{-1}(\nabla\rho)+{\cal O}(\lambda^{1/2})$
and $f_0=f+{\cal O}(\lambda^{1/2})$, that
\begin{eqnarray}\label{E2-form2}
   E_2(x) & = & 4\pi\Delta^{-1}(\nabla\rho_2)(x)
   -2\pi\int\int\frac{1}{|x-y|}\,
   \Delta^{-1}(\nabla\rho)(y)\,f(y, p)\,dy\,dp \nonumber
   \\ & & -\,2\pi\int\int\frac{x-y}{|x-y|^3}\,(x-y)\cdot\Delta^{-1}(\nabla\rho)(y)
   \,f(y, p)\,dy\,dp \nonumber
   \\ & & +\,\frac{1}{2}\int\int\bigg(\frac{p^2}{|x-y|^3}
   -\frac{3(p\cdot(x-y))^2}{|x-y|^5}\bigg)(x-y)\,f(y, p)\,dy\,dp
   \nonumber \\ & & +\,{\cal O}(\lambda^{1/2}).
\end{eqnarray}
In (\ref{Lor-force}) the field $E_2$ carries the factor $\lambda$.
Whence we can as well use $E_2$ in its form (\ref{E2-form2}) rather
than (\ref{E2-form}) to obtain a valid approximation of $F_L$
up to order ${\cal O}(\lambda)$. In particular, (\ref{E2-form2})
is more convenient to compare this approximation to what comes out
from the associated particle model. In \cite{MK-S} it was show
that the dynamics of $N$ particles coupled to their self-generated Maxwell field
can be described over long times by means of the effective equations
\begin{equation}\label{max-lor}
   \frac{d}{dt}\,T_{{\rm kin}}(v_\alpha)=G_\alpha^{{\rm CD}}(q, v, \dot{v})
   +{\cal O}(\lambda^{3/2}),\quad 1\le\alpha\le N,
\end{equation}
where $T_{{\rm kin}}$ is the kinetic energy, and the force $G_\alpha^{{\rm CD}}$
is given by
\begin{eqnarray}\label{G-form}
   \lefteqn{G_\alpha^{{\rm CD}}(q, v, \dot{v})} \nonumber \\ & = &
   e_\alpha\sum_{\stackrel{\beta=1}{\beta\neq\alpha}}^N
   e_\beta\,\frac{\xi_{\alpha\beta}}{|\xi_{\alpha\beta}|^3} \nonumber
   +\lambda e_\alpha\sum_{\stackrel{\beta=1}{\beta\neq\alpha}}^N
   e_\beta\Bigg(-\frac{1}{2|\xi_{\alpha\beta}|}\,\dot{v}_\beta
   -\frac{(\dot{v}_\beta\cdot \xi_{\alpha\beta})}
   {2|\xi_{\alpha\beta}|^3}\,\xi_{\alpha\beta} \nonumber \\ & &
   +\,\frac{v_\beta^2}{2|\xi_{\alpha\beta}|^3}\,\xi_{\alpha\beta}
   -\frac{3{(v_\beta\cdot\xi_{\alpha\beta})}^2}{2|\xi_{\alpha\beta}|^5}
   \,\xi_{\alpha\beta}-\frac{(v_\alpha\cdot v_\beta)}{|\xi_{\alpha\beta}|^3}\,
   \xi_{\alpha\beta}+\frac{(v_\alpha\cdot\xi_{\alpha\beta})}
   {|\xi_{\alpha\beta}|^3}\,v_\beta\Bigg).\qquad
\end{eqnarray}
Here particle $\alpha$ has position $q_\alpha(t)\in\R^3$ and velocity
$v_\alpha(t)=\dot{q}_\alpha(t)$. Moreover, $\xi_{\alpha\beta}=q_\alpha-q_\beta$.
Note that in \cite[Lemma 3.5]{MK-S}, eq.~(\ref{max-lor}) is formulated
on a different scale by means of a dimensionless parameter $\epsilon\ll 1$.
In this units, the error was ${\cal O}(\epsilon^{7/2})$, and to pass to (\ref{max-lor})
it is first necessary to multiply by $\epsilon^{-2}$ to undo the transformation,
and then $\epsilon\cong\lambda$, whence the error term
is of order ${\cal O}(\lambda^{3/2})$. In addition, in \cite{MK-S}
the charge distribution $\rho$ and the current $j$ in the Maxwell
equations are considered without the factor $4\pi$. Hence to adjust
to (\ref{E-eq}), (\ref{B-eq}) the expression for $G_\alpha^{{\rm CD}}$
had to be multiplied by $4\pi$. We also remark that (\ref{max-lor})
holds in the limit of slow particles which are far apart, cf.~\cite{MK-S}.
The equation $\frac{d}{dt}\,T_{{\rm kin}}(v_\alpha)=G_\alpha^{{\rm CD}}(q, v, \dot{v})$
is governed by a suitable Lagrangian, consisting of a Coulomb
and a Darwin contribution, cf.~\cite[Ch.~12.6]{jackson} and \cite[Eq.~(1.1)]{MK-S}.
Since the Abraham model considered in \cite{MK-S} is only semi-relativistic,
the particular form of $T_{{\rm kin}}$ cannot be expected to agree
with the corresponding expression resulting from the Vlasov-Maxwell system
through expansion in powers of $\lambda^{1/2}$; see \cite{apkies} for a recent
improved model. However, in the formal limit $N\to\infty$
the force $G_\alpha^{{\rm CD}}$ coincides with $F_L\cong E_0+\lambda^{1/2}E_1
+\lambda (E_2+v\times B_1)$, using $E_2$ from (\ref{E2-form2}). To see this,
we fix one particle $\alpha$, i.e., its position $x\cong q_\alpha\in\R^3$
and its velocity $v\cong v_\alpha\in\R^3$. Setting $e_\beta=1$ for
all $\beta$ corresponding to (\ref{vlas}), then
\[ \sum_{\stackrel{\beta=1}{\beta\neq\alpha}}^N
   \frac{\xi_{\alpha\beta}}{|\xi_{\alpha\beta}|^3}
   \stackrel{N\to\infty}{\longrightarrow}
   \int\int\frac{x-y}{|x-y|^3}f(x, p)\,dy\,dp
   =4\pi\Delta^{-1}(\nabla\rho)(x) \]
results in the Coulomb part; note that the corresponding expression in $F_L$
is given by $E_0+\lambda^{1/2}E_1+\lambda 4\pi\Delta^{-1}(\nabla\rho_2)
=4\pi\Delta^{-1}(\nabla\rho)+{\cal O}(\lambda^{3/2})$. In addition, we obtain
\begin{eqnarray*}
   \lefteqn{\lambda\sum_{\stackrel{\beta=1}{\beta\neq\alpha}}^N
   \bigg(-\frac{(v_\alpha\cdot v_\beta)}{|\xi_{\alpha\beta}|^3}\,
   \xi_{\alpha\beta}+\frac{(v_\alpha\cdot\xi_{\alpha\beta})}
   {|\xi_{\alpha\beta}|^3}\,v_\beta\Bigg)} \\ & \stackrel{N\to\infty}{\longrightarrow} &
   \lambda\int\int\frac{dy\,dp}{|x-y|^3}
   \,v\times (p\times (x-y))\,f(x, p)=\lambda\,v\times B_1(x)+{\cal O}(\lambda^{3/2}),
\end{eqnarray*}
cf.~(\ref{B1-form}). To deal with the remaining part of (\ref{G-form})
we have to reexpress the $\dot{v}_\beta$ through lower order terms.
Since $\lambda(-\frac{1}{2|\xi_{\alpha\beta}|}\,\dot{v}_\beta
-\frac{(\dot{v}_\beta\cdot \xi_{\alpha\beta})}{2|\xi_{\alpha\beta}|^3}
\,\xi_{\alpha\beta})={\cal O}(\lambda)$, we can simply substitute
\[ \dot{v}_\beta=m_\beta\dot{v}_\beta
   =\sum_{\stackrel{\nu=1}{\nu\neq\beta}}^N
   \frac{\xi_{\beta\nu}}{|\xi_{\beta\nu}|^3}+{\cal O}(\lambda)
   \stackrel{N\to\infty}{\longrightarrow}4\pi\Delta^{-1}(\nabla\rho)(y)
   +{\cal O}(\lambda) \]
without changing the validity of (\ref{max-lor}). Hence it follows that
\begin{eqnarray*}
  \lefteqn{\lambda\sum_{\stackrel{\beta=1}{\beta\neq\alpha}}^N
   \Bigg(-\frac{1}{2|\xi_{\alpha\beta}|}\,\dot{v}_\beta
   -\frac{(\dot{v}_\beta\cdot \xi_{\alpha\beta})}
   {2|\xi_{\alpha\beta}|^3}\,\xi_{\alpha\beta}
   +\frac{v_\beta^2}{2|\xi_{\alpha\beta}|^3}\,\xi_{\alpha\beta}
   -\frac{3{(v_\beta\cdot\xi_{\alpha\beta})}^2}{2|\xi_{\alpha\beta}|^5}
   \,\xi_{\alpha\beta}\Bigg)} \\ & & \stackrel{N\to\infty}{\longrightarrow}
   \lambda\int\int\bigg(-\frac{2\pi}{|x-y|}\,\Delta^{-1}(\nabla\rho)(y)
   -2\pi\frac{x-y}{|x-y|^3}\,\Delta^{-1}(\nabla\rho)(y)\cdot (x-y)
   \\ & & +\,\frac{p^2}{2|x-y|^3}\,(x-y)-\frac{3{(p\cdot (x-y))}^2}{2|x-y|^5}
   \,(x-y)\Bigg)\,f(y, p)\,dy\,dp+{\cal O}(\lambda^{3/2}),
\end{eqnarray*}
in agreement with (\ref{E2-form2}).

To summarize, up to the Darwin order the expansion in powers
$\lambda^{1/2}$ of the Vlasov-Maxwell system does agree
with the infinite particle number limit of the particle system's
effective dynamics.

\subsection{Approximation to the order of radiation reaction}

The purpose of this section is to push the expansion of the Lorentz force
one step further to include the terms of order ${\cal O}(\lambda^{3/2})$,
i.e., we consider
\begin{equation}\label{Lor-force-2}
   F_L=E_0+\lambda^{1/2}E_1+\lambda (E_2+v\times B_1)
   +\lambda^{3/2}(E_3+v\times B_2)+{\cal O}(\lambda^2),
\end{equation}
cf.~(\ref{Lor-force}). The equations to be solved for the coefficients
$E_3$ and $B_2$ are $\div E_3=4\pi\rho_3$, $\curl E_3=-\partial_t B_2$,
$\div B_2=0$, and $\curl B_2=\partial_t E_1+4\pi j_1$, and this leads to
\[ \quad\Delta E_3=4\pi\nabla\rho_3
   +\partial_t(\partial_t E_1+4\pi j_1),\quad\Delta B_2
   =-4\pi\,\curl j_1. \]
The solution of the latter equation is analogous to that of (\ref{B1-form}) 
and just
contributes to the expansion of $j$. Concerning $E_3$, it could be treated
just as $E_2$ was, obtaining a solution which vanished at infinity. There
would be no trace of radiation reaction and this solution is not the
appropriate one for obtaining an approximation of a solution of the Maxwell
equations at the 1.5PN level. The equations of the pure post-Newtonian
expansion of section 2 do not contain information about damping in this
case. They need to be supplemented by asymptotic conditions different from
the condition of vanishing at infinity. In the case of the Maxwell equations
this additional information could be obtained by doing expansions of the
fundamental solution different from the post-Newtonian one.

As in the previous section \ref{CD-sect}, the connection can be made between
this result and the $N\to\infty$ limit of a system of $N$ particles.
In \cite{MK-S-2} it was proved that to the order of radiation reaction
the particle dynamics is governed over long times by
\[ \frac{d}{dt}\,T_{{\rm kin}}(v_\alpha)=G_\alpha^{{\rm CD}}(q, v, \dot{v})
   +G_\alpha^{{\rm RR}}(q, v, \dot{v})+{\cal O}(\lambda^2),\quad 1\le\alpha\le N, \]
cf.~(\ref{max-lor}), with the new ${\cal O}(\lambda^{3/2})$-term
\begin{eqnarray}\label{GRR-def}
   G_\alpha^{{\rm RR}}(q, v, \dot{v}) & = & \frac{e_\alpha}{12\pi}\lambda^{3/2}
   \sum_{\stackrel{\beta, \beta'=1}{\beta\neq\beta'}}^N e_\beta e_{\beta'}
   \bigg(\frac{e_\beta}{m_\beta}-\frac{e_{\beta'}}{m_{\beta'}}\bigg)
   \bigg(\frac{1}{|q_{\beta}-q_{\beta'}|^3}(v_\beta-v_{\beta'})
   \nonumber \\ & & \hspace{8em}
   -\,\frac{3(q_{\beta}-q_{\beta'})\cdot (v_\beta-v_{\beta'})}
   {|q_{\beta}-q_{\beta'}|^5}\,(q_{\beta}-q_{\beta'})\bigg)\qquad
\end{eqnarray}
accounting for the radiation reaction. We do assume now that there are two species
of particles of equal unit mass and of total number $N=2M$, such that $e_\alpha=e=1$
for $\alpha=1, \ldots, M$ and $e_\alpha=-e=-1$ for $\alpha=M+1, \ldots, 2M$.
The respective particle number densities are denoted by $f^+$ and $f^-$. Decomposing the
$\sum_{\beta, \beta'=1, \beta\neq\beta'}^N$ in (\ref{GRR-def}) accordingly,
we see that
\begin{eqnarray}\label{GRR-lim}
   G_\alpha^{{\rm RR}}(q, v, \dot{v}) & \stackrel{N\to\infty}{\longrightarrow} &
   \frac{1}{3\pi}\lambda^3\,\int\int dx_1\,dp_1\,\int\int dx_2\,dp_2\,
   f^+(x_1, p_1)f^-(x_2, p_2) \nonumber \\ & & \hspace{-3em}
   \cdot\bigg(\frac{1}{|x_1-x_2|^3}(p_1-p_2)
   -\frac{3(x_1-x_2)\cdot (p_1-p_2)}{|x_1-x_2|^5}\,(x_1-x_2)\bigg).\qquad
\end{eqnarray}
On the other hand, recalling the dipole moment $D(t)$ from (\ref{D-def})
and $D^{[2]}(t)$ from (\ref{D2-def}) we have
\[ \dot{D}(t)=\int\int p\,(f^+ -f^-)\,dx\,dp,
   \quad\ddot{D}(t)=D^{[2]}(t)+{\cal O}(\epsilon). \]
Differentiating the latter relation w.r.t.~time, and using the explicit form
of the electric field $\nabla U$ obtained from $\Delta U=4\pi\rho$
in conjunction with the Vlasov equations (\ref{vlas+}) and (\ref{vlas-})
for $f^+$ and $f^-$, we obtain through explicit evaluation that
\begin{eqnarray*}
   \lefteqn{\stackrel{...}{D}(t)} \\ & = & \int\int\int\int\,
   dx_1\,dp_1\,dx_2\,dp_2\,(p_2\cdot\nabla_{x_2})\frac{x_1-x_2}{|x_1-x_2|^3}
   \\ & & \hspace{5em} \cdot(f^+(x_2, p_2)-f^-(x_2, p_2))
   \,(f^+(x_1, p_1)+f^-(x_1, p_1))
   \\ & & +\,\int\int\,dx_1\,dp_1\,(p_1\cdot\nabla)\nabla U\,
   (f^+ +f^-)+{\cal O}(\epsilon) \\ & = & \int\int\int\int\,
   dx_1\,dp_1\,dx_2\,dp_2\,\bigg(\frac{1}{|x_1-x_2|^3}(p_1-p_2)
   \\ & & \hspace{13em} -\,\frac{3(x_1-x_2)\cdot (p_1-p_2)}{|x_1-x_2|^5}\,(x_1-x_2)\bigg)
   \\ & & \hspace{5em} \cdot(f^+(x_2, p_2)-f^-(x_2, p_2))
   \,(f^+(x_1, p_1)+f^-(x_1, p_1))+{\cal O}(\epsilon).
\end{eqnarray*}
Now we observe that using the change of variables
$x_1\leftrightarrow x_2$ and $p_1\leftrightarrow p_2$ it is clear that
the term containing $f^+(x_2, p_2)f^+(x_1, p_1)$ as well as the one containing
$f^-(x_2, p_2)f^-(x_1, p_1)$ does vanish, whereas the other two terms are the same
and add up. Consequently we finally find
\begin{eqnarray*}
   \stackrel{...}{D}(t) & = & 2\int\int\int\int\,dx_1\,dp_1\,dx_2\,dp_2\,
   \bigg(\frac{1}{|x_1-x_2|^3}(p_1-p_2)
   \\ & & \hspace{4em} -\,\frac{3(x_1-x_2)\cdot (p_1-p_2)}{|x_1-x_2|^5}\,(x_1-x_2)\bigg)
   f^+(x_1, p_1)f^-(x_2, p_2).
\end{eqnarray*}
Comparing this result to (\ref{GRR-lim}) we see that the VPD model
considered in section \ref{hyb-sect} is a quite reasonable one
to describe electromagnetic radiation reaction.

\section{Outlook}

In this paper various models of radiation damping and their mutual
relations have been discussed. Potential difficulties in establishing
mathematical results about these models have been pointed out. In the
models of section 1 the force on a particle is determined instantaneously
by the distribution of all particles. These were self-contained models
of continuum mechanics. Another type of model for radiation damping has
been studied in the literature. In that case the description of matter
is a schematic one (one or more oscillators) while the field satisfies
a wave equation. Examples are a model of Burke \cite{burke} and one
of Aichelburg and Beig \cite{aichelburg} whose mathematical properties
were further studied in \cite{hoenselaers} and \cite{stewart}. It would
be interesting to make connections between the latter models and those
discussed in the present paper.

The central open question concerning the models for electromagnetic
radiation damping in section 2 is that of
proving that solutions of these models approximate solutions of the
Vlasov-Maxwell system in some appropriate sense. The results of section 3
may help to give indications how results of this kind should be
formulated and proved. However all that was done up to now was to point
out a formal similarity between the equations. If a complete rigorous
analysis of the case of the Vlasov-Maxwell system (or even of a model
such as that of \cite{burke}) were obtained then it could serve as a basis
for an attack on the much harder case of radiation damping for the Einstein
equations. \smallskip

\noindent {\bf Acknowledgements:} The authors are grateful to
H.~Spohn, M.~Kiessling and B.~Schmidt for discussions.

\end{document}